\documentclass{article}

\usepackage[margin=2cm]{geometry}

\usepackage{amsmath}
\usepackage{xcolor}
\usepackage{booktabs}
\usepackage{lineno,hyperref,geometry}
\usepackage{soul}
\modulolinenumbers[2]
\newcommand{\ra}[1]{\renewcommand{\arraystretch}{#1}}

\usepackage{siunitx}

\providecommand{\keywords}[1]{\noindent\textbf{ Keywords:} #1}
\usepackage{authblk}
\usepackage{graphicx}
\usepackage[maxnames=5,style=numeric-comp,firstinits=true,sorting=none,backend=bibtex]{biblatex}
\renewbibmacro{in:}{}
\begin{document}

\title{Comparison of Multiobjective Optimization Methods for the LCLS-II Photoinjector}

\author[1]{Nicole Neveu}

\author[2]{Tyler H. Chang}
\author[1,3]{Paris Franz}

\author[2]{Stephen Hudson}

\author[2]{Jeffrey Larson}

\affil[1]{Accelerator Directorate, SLAC National Accelerator Laboratory \linebreak[4] \texttt{\small nneveu@slac.stanford.edu}; \texttt{\small paris@slac.stanford.edu}}
\affil[2]{Mathematics and Computer Science Division, Argonne National Laboratory \linebreak[4] \texttt{\small tchang@anl.gov}; \texttt{\small shudson@anl.gov}; \texttt{\small jmlarson@anl.gov}}
\affil[3]{Department of Applied Physics, Stanford University}

\maketitle

\begin{abstract}
Particle accelerators are among some of the largest science experiments in the world
and can consist of thousands of components with a wide variety of input ranges.
These systems can easily become unwieldy optimization problems during design and operations studies.
Starting in the early 2000s,  searching for better beam dynamics configurations became synonymous with
heuristic optimization methods in the accelerator physics community.
Genetic algorithms and particle swarm optimization are currently the most widely used.
These algorithms can take thousands of simulation evaluations to find optimal solutions for one machine prototype.
For large facilities such as the Linac Coherent Light Source (LCLS) and others,
this equates to a limited exploration of many possible design configurations.
In this paper, the LCLS-II photoinjector is optimized with three optimization algorithms.
All optimizations were started from both a uniform random and Latin hypercube sample.
In all cases, the optimizations started from Latin hypercube samples outperformed optimizations started from uniform samples.
All three algorithms were able to optimize the photoinjector, with the
model-based methods approximating the Pareto
front in fewer simulation evaluations. This work, in combination with previous
optimization observations, indicates objective penalties have a strong impact
on the efficiency of such methods. In general, we recommend heuristic methods
for initial optimizations and model-based methods when information
about the objective space is available.
\end{abstract}


\section{Motivation}
Particle accelerators are large mechanical and electrical systems that are inherently difficult to optimize given the number of independent parameters.
Colliders, free-electron lasers (FELs), ultrafast electron diffraction (UED) facilities, and other accelerator complexes range in length from a few meters to a few kilometers. The largest accelerator in the world, at CERN, has a circumference of 27 km.
Accelerators on the kilometer scale and below consist of thousands of components with variable inputs.
Because of the scale and computational modeling constraints, optimization of accelerators is rarely done as a whole~\cite{QIANG2022166294}.
Sections of the machine are optimized in a piecewise fashion, typically delineated by beam energy.
Even with a bite-size approach, the number of possible optimization variables ranges from tens to hundreds of variables.
This often motivates researchers to consider simplified models of the entire system.

The overwhelming parameter space may indicate why heuristic optimization methods have gained popularity in the past two decades, with genetic algorithms (GAs) and particle swarm optimization being two of the most widely used methods.
GAs have become particularly widespread, with implementations in several accelerator physics simulation codes.
While effective, these algorithms can require large computational resources depending on the problem.
For large facilities such as the LCLS,
this equates to many design configurations not being fully explored.
At present, a typical NSGA-II optimization of the LCLS-II photoinjector
can require thousands of core-hours per configuration (using low-fidelity simulations).
If a key static parameter changes---for example, the bunch charge or a magnet location---the optimization process must be repeated.
As a result, millions of core-hours are spent yearly at high-performance computing (HPC)
facilities such as the National Energy Research Scientific Computing Center.
While there is nothing inherently wrong with this scenario,
improvement in optimization efficiency for these problems could open several doors for the accelerator physics community.
A shorter time to solution (i.e., fewer simulation evaluations) would allow for wider searches in the design and experiment planning phases.
Fewer required computing resources would also allow students and small facilities or schools to perform optimization studies
without the need to apply for or request computing resources at an HPC facility.

In this paper, optimization of the LCLS-II photoinjector is used to compare three optimization methods. This is an established optimization problem that is actively being solved for design and commissioning studies.
HPC resources are currently required to solve this problem with standard methods.
It is also a typical optimization case (a photoinjector) that is broadly applicable to machines at other accelerator labs, including
FELs, UED facilities, energy recovery linacs, and medical linacs,
since all these R\&D facilities use electron injectors of some kind.
In Section~\ref{sec:opt} we describe the optimization methods used.
Section~\ref{sec:phys} gives a detailed description of the optimization problem.
A comparison of the optimization results and the difference
in simulation evaluations needed in each case (convergence) is shown in Section~\ref{sec:results}.
The results are discussed in Section~\ref{sec:discussion}.
All codes and resources needed to reproduce these results are presented in Section~\ref{sec:code}.

\section{Optimization methods}\label{sec:opt}
Three optimization methods are compared in this paper: scalarization, a genetic algorithm, and response surface (i.e., surrogate-based)  methodology.
To aid in the following sections, we define the general terms in the optimization problem here.
Each evaluation requires running a simulation code given twelve input parameters, which will be denoted as $\vec{v}$.
The function being minimized will be referenced as $f\left(\vec{v}\right)$ or $f\left(\vec{v}, w\right)$ depending on the optimization method, where $w$ represents a weight applied to objective values.
Each optimization case was given boundaries on $\vec{v}$ and identical initial samples.
Values in $\vec{v}$ and objectives in each case are scaled during optimization such that they are on the same order of magnitude.
This is done to resolve differences in units that would mislead the optimization algorithms.
For example, the input laser radius has units of millimeters as opposed to the laser's full width at half maximum, which is in picoseconds.
Along with the same search space and scaling, each method was allowed to evaluate the same number of simulations.
In this way, convergence can be compared fairly with respect to the same budget of evaluations.

\subsection{Genetic algorithm: NSGA-II}
GAs are a class of evolutionary optimization methods based on gene mutations in nature.
The nondominated sorted genetic algorithm-II~(NSGA-II), described by K.~Deb~et~al.~in~\cite{dpam:02},
is the most widely used optimization algorithm in the accelerator physics community.
Derived from~\cite{dpam:02}, this algorithm can roughly be described in six steps:
\begin{enumerate}
    \item Supply or perform initial sample with $N$ evaluations. This is the parent population, $P_0$.
    \item Apply nondominated sorting to assign a nondomination rank to all points in the current population $P_i$.~\label{step:two}
    \item Calculate a crowding distance score for all members of $P_i$.~\label{step:three}
    \item Use binary tournament selection (using the nondominated rank and crowding distance), recombination, and mutation operators on $P_i$ to generate a batch of children ($V_i$) of configurable size. In NSGA-II, this size parameter is known as the {\it population size}.~\label{step:four}
    \item Evaluate all objectives for all children $\vec{v}\in V_i$, and combine this set with the entire parent population~$P_{i}$ to generate the next population:
    $P_{i+1} = P_i \cup V_i$.
    \item Start repetition at Step~\ref{step:two}.
\end{enumerate}
Step~\ref{step:four} is the part of the algorithm most resembling nature, where individuals ($\vec{v}\in P_i$) in the population are selected for mutation or not.
Binary tournament selection refers to the comparison of two randomly picked individuals in the population; the individual with the better objective values is chosen for recombination/crossover.
In NSGA-II, this is based on a unique partial ordering operator, which combines the
nondominated rank and crowding distance of each point to produce a single
fitness score.
In recombination, values from two individuals that survived the tournament will be used to generate a child point.
When a point is chosen for mutation, values in $\vec{v}$ can change randomly by a small amount.
Further explanation of Step~\ref{step:four} can be found in the original paper~\cite{dpam:02}.

The following makes NSGA-II unique compared with other multiobjective GAs:
\begin{itemize}
\item NSGA-II is an {\it elitist} algorithm, meaning that all nondominated points in the current population are maintained in future populations and returned in the final solution.
Therefore, it is not possible for NSGA-II to ``regress'' by dropping nondominated members of the current population without generating suitable replacements among their batch of children.
\item NSGA-II generates batches of children that are (heuristically) well distributed over the current space of nondominated points, because of its usage of the crowding distance metric during Step~\ref{step:three}.
This allows NSGA-II to produce a good spread of solutions without additional hyperparameters.
\item To maintain elitism over large population sizes, without becoming computationally intractable, NSGA-II employed a novel (at the time) ``fast nondominated sorting'' algorithm.
\end{itemize}

Further, evolutionary methods such as NSGA-II are designed to
(one hopes)
identify better local optima because of randomization. Unfortunately, such
methods can require many simulation evaluations, namely,
dozens of generations, each requiring many dozens of individual simulation
evaluations. While these evaluations can be performed concurrently, relying on
randomization to find progress may still result in unsatisfactory runtimes.
Regardless, considerable effort has been invested in adapting NSGA-II to a variety of accelerator physics problems~\cite{bazarov05,hofler13,gull1,gull2,marija}. Several commonly used accelerator simulation codes have incorporated this algorithm for use by users directly or into their distributions, or provide a separate but paired package for optimization.
Examples include OPAL~\cite{opal} (the simulation code used in this paper),
BMAD~\cite{Sagan:Bmad2006}, Elegant~\cite{elegant}, and General Particle Tracer~\cite{gpt}.

One advantage, and possibly a key reason this algorithm has been widely used in the accelerator physics community,
is the ease of multiobjective setup (given an implementation is already written and provided).
Since many implementations and examples exist within the accelerator physics community,
users, to some degree, can provide two objectives, set the boundaries, and hit go.
This statement, however, ignores the considerable complexity of hyperparameter tuning, constraint options, and the impact of boundaries on convergence. Often hyperparameter tuning and constraint or boundary evaluations are
forgone for deadline constraints and the stresses of running an accelerator facility.

\subsection{Multistart with local method (scalarization)}
As functions of the beamline parameters, typical objectives such as the emittance and the bunch length
often display many local optima. Therefore, local optimization runs (regardless of the
method) can produce solutions with qualities that greatly vary depending on
how the run is initialized. This is one of the most repeated arguments against using local optimization methods for accelerator physics problems, yet this point usually ignores the multistart approach.

Optimization methods that build and use (local) models of simulation output
as a function of the input parameters typically require significantly fewer
evaluations to identify local optima. BOBYQA~\cite{Powell2009a} is one such
popular method that builds local quadratic models of an objective function near
a best-known point. The quadratic model is then optimized in a neighborhood of
the best-known point to produce a candidate point. If the candidate is better,
it becomes the best-known point; otherwise, the model or neighborhood needs to be
updated. BOBYQA is used here as a local optimization method in our numerical
tests.

As many would point out, the performance of BOBYQA is especially sensitive to the starting point used,
so it is incorporated within a multistart framework. The benefits of using a
local optimization method in order to use fewer simulation calls can be erased
if we start too many local optimization runs. Therefore, we use the
APOSMM~\cite{Larson2018} algorithm to decide when and where to start a local
optimization run. APOSMM starts with a (coarse) sampling of the parameter space
and starts runs from points that do not have better points nearby. (Nearby is
defined by a radius that adjusts as the algorithm progresses.)

\subsection{VTMOP (surrogate-based method)}
VTMOP is a solver for computationally expensive black-box multiobjective
optimization problems, which combines surrogate modeling, trust-region methods,
direct search techniques, and an adaptive weighted-sum scalarization scheme.
VTMOP is based on an adaptive weighting scheme for sampling the Pareto front,
described in \cite{deshpande2016multiobjective};
its implementation is fully described by
Chang et al.~\cite{chang2022algorithm}.
The algorithm can be summarized by the following four-step iterative process:

\begin{enumerate}
\item
If this is not the first iteration, identify an isolated point in the
current approximation to the Pareto front.
Fit a local trust region (LTR) centered at the corresponding design point,
and choose several adaptive weight vectors.
If this is the first iteration, then there is no current nondominated set;
the entire design space must be explored, and the initial weight vectors
are predetermined.

\item
Perform an exploration by taking evaluations throughout the current LTR
(or the entire design space if this is the first iteration).

\item
Fit a surrogate to each component of the multiobjective cost function/simulation
output,
using the data gathered from Step 2 (plus any data available from previous
iterations).
Apply each weight vector to the surrogates, and solve the resulting
scalarized surrogate problems by using a direct search strategy.
The solutions to these surrogate problems form a batch of candidate design
points.

\item
Evaluate every point in the current batch; and if the budget is not exceeded,
proceed to the next iteration.
\end{enumerate}

Modifications to the original algorithm were made in order to take advantage
of parallel computing resources.
The main modification involved padding out the batch of candidate designs
produced in Step 3 to match the available computing resources.
The modifications are fully described in~\cite{chang2020managing}.

Note that VTMOP has several algorithm parameters, such as the trust-region
radius, starting search budget, subsequent search budgets, trust-region decay
factor, and minimum trust-region radius, which could potentially be tuned to
improve its performance.
Since the default values for these parameters were carefully chosen
to provide reasonable performance on a wide variety of problems  and since our
readers are unlikely to tune VTMOP themselves, we have elected to use VTMOP's
default parameters.
These algorithm parameters and their default values are fully described in \cite{chang2022algorithm}.

\section{Optimization of the LCLS-II photoinjector}\label{sec:phys}

\begin{figure*}[h]
	\centering
	\includegraphics[width=\linewidth]{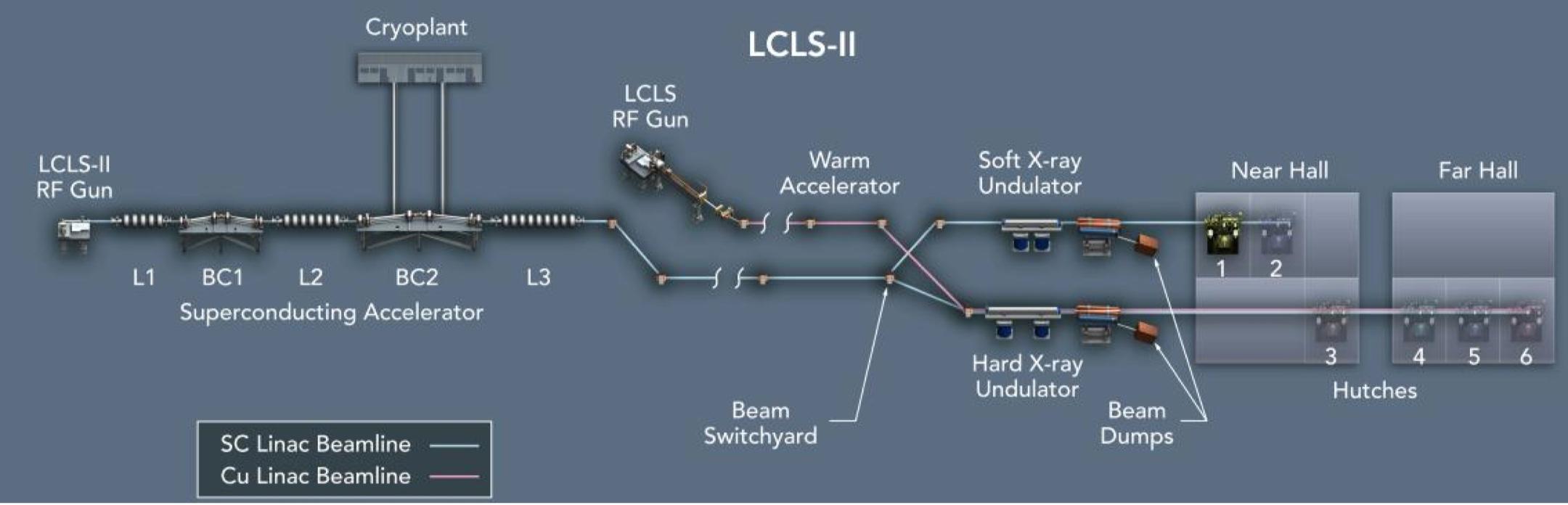}
	\caption{Layout of LCLS copper and superconducting accelerators and undulator halls.
  \label{fig:lcls_layout}}
\end{figure*}

SLAC has been home to a variety of linear accelerators since the 1960s~\cite{slac50}.
With the advent of LCLS, the design and operation of photoinjectors have become an area of focus for the lab.
LCLS is being upgraded (LCLS-II), for which construction of a superconducting injector and linear accelerator (linac) was recently completed and commissioning began in 2022.
LCLS-II shares the same tunnel housing that the copper LCLS injector is operating in.
Both photoinjectors will supply electron bunches to the undulator hall,
with the lines being interchangeable, as shown in Fig.~\ref{fig:lcls_layout}.
In this section, details are given on the input variables, boundary conditions, physics objectives, and beamline components.

\subsection{Design variables}~\label{sec:dvar}
\begin{figure*}[h]
    \centering
    \includegraphics[width=0.31\linewidth]{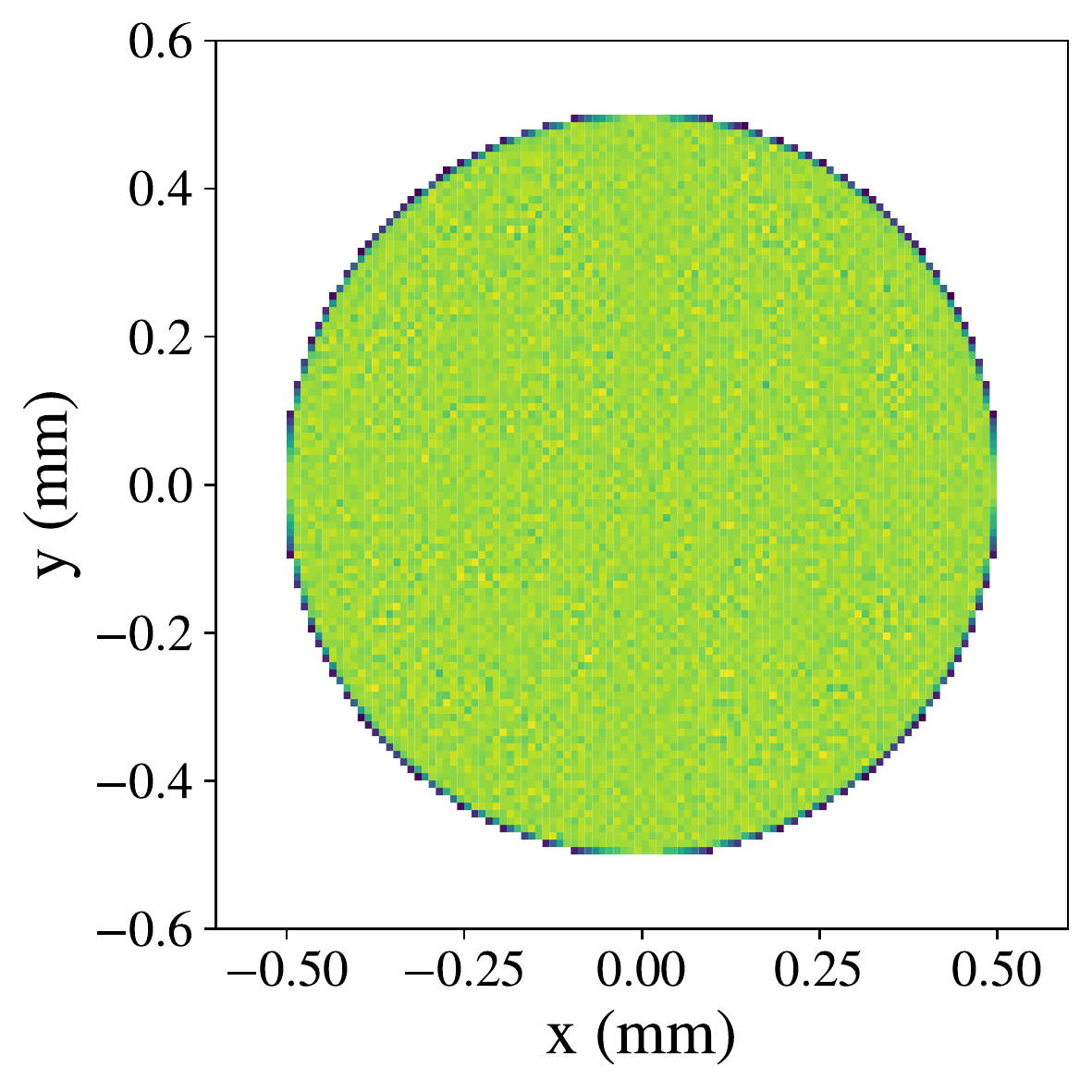}
    \includegraphics[width=0.46\linewidth]{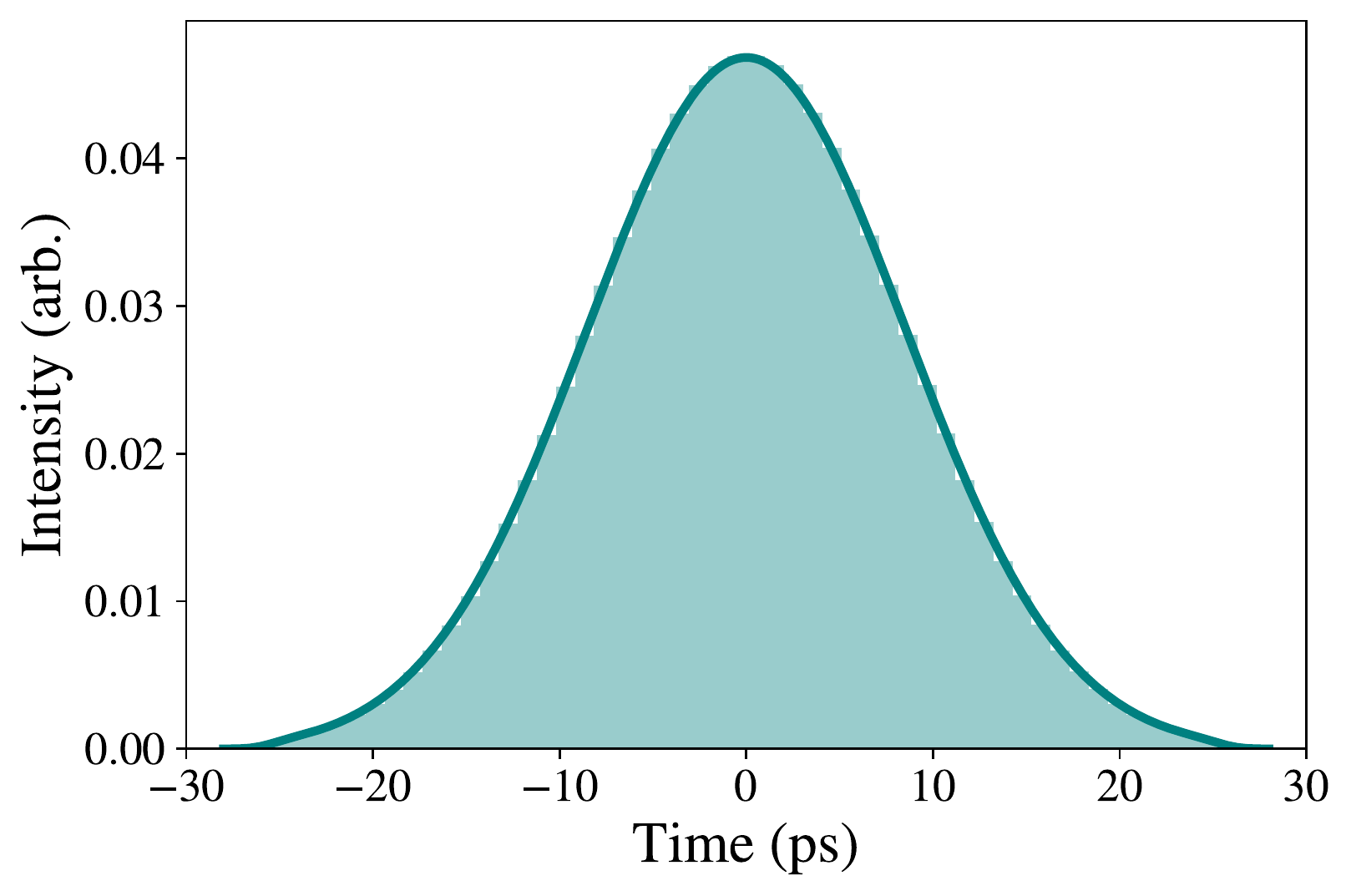}
    \caption{Typical transverse (left) and longitudinal (right) laser profiles. The radius and longitudinal FWHM are included as optimization variables in the studies shown here. \label{fig:laser}}
\end{figure*}

The new LCLS superconducting linac will be fed by a high-repetition-rate
quarter-cell photoinjector (rf gun), operating at~\SI{187}{MHz}.
This rf gun is a replica of the APEX~\cite{apex} gun
that was developed at Lawrence Berkeley National Laboratory
and will allow LCLS-II to operate in continuous wave mode with an electron repetition rate of~\SI{1}{MHz}.
This is a major improvement over the previous repetition rate of~\SI{120}{Hz},
since it allows for an increase in the x-ray pulse delivery frequency by over 8,000 times.
Input variables in this section include parameters related to the laser, rf phases, and magnet settings.
The laser used to generate electrons at the cesium telluride cathode
is a Class-IV infrared laser operated at \SI{1030}{nm}.
The fourth harmonic of the laser is taken to generate \SI{257.5}{nm} ultraviolet pulses.
At the time of this study, efforts were ongoing to determine the best possible laser shape, given operational constraints and commissioning goals.
The transverse laser radius and longitudinal full width at half maximum (FWHM), are included as optimization variables.
For each simulation, a new laser distribution is generated and fed to the beam dynamics code.
An example laser distribution in both the transverse and longitudinal dimensions is shown in Fig.~\ref{fig:laser}.

\begin{figure*}[h]
    \centering
    \includegraphics[width=0.75\linewidth]{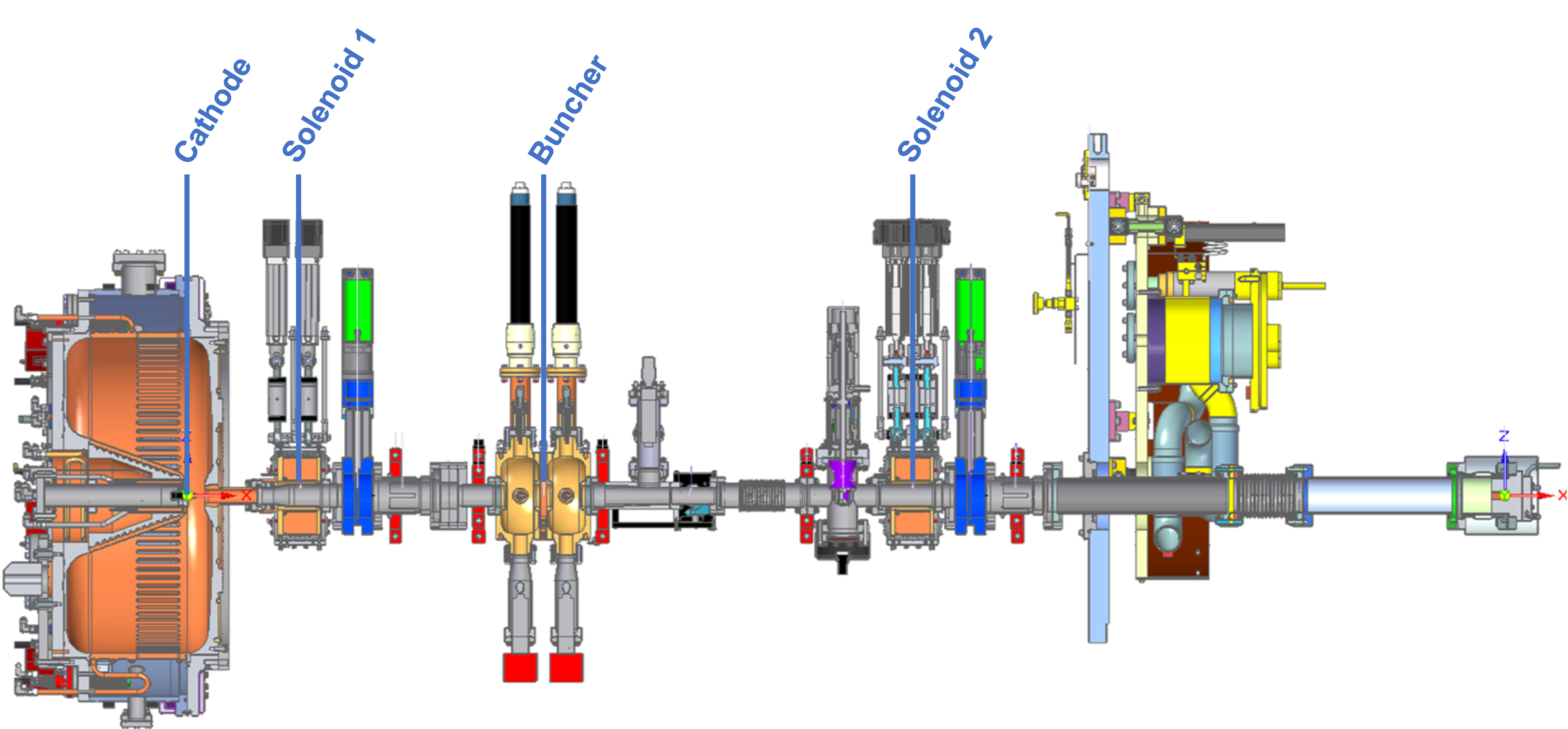}
    \caption{LCLS-II photoinjector that feeds into the LCLS-II superconducting linear accelerator.
        The laser parameters, gun phase, and solenoid strengths are included in the optimization variables.
        This hardware is currently installed and has been operated during commissioning.
    \label{fig:gun}}
\end{figure*}

The next optimization variable is the phase in the rf gun,
which indicates the arrival time of the laser pulse with respect to the rf pulse in the gun cavity.
Directly following the gun is the first focusing magnet (solenoid), then a buncher cavity.
The buncher cavity is used to reduce the bunch length and/or increase the energy depending on the
phase and gradient values chosen by the optimizer. The phase for the buncher indicates the beam arrival time with respect to the rf field,
and the gradient refers to the peak amplitude of the rf field in the cavity. Larger gradients result in higher beam energies at this location.
The last variable in this section is the next solenoid, which has the same bounds and operating regime as the first.
Figure~\ref{fig:gun} includes the rf gun, buncher, and solenoid layout as currently installed and operated at SLAC.

\begin{figure*}[h]
	\centering
	\includegraphics[width=0.75\linewidth]{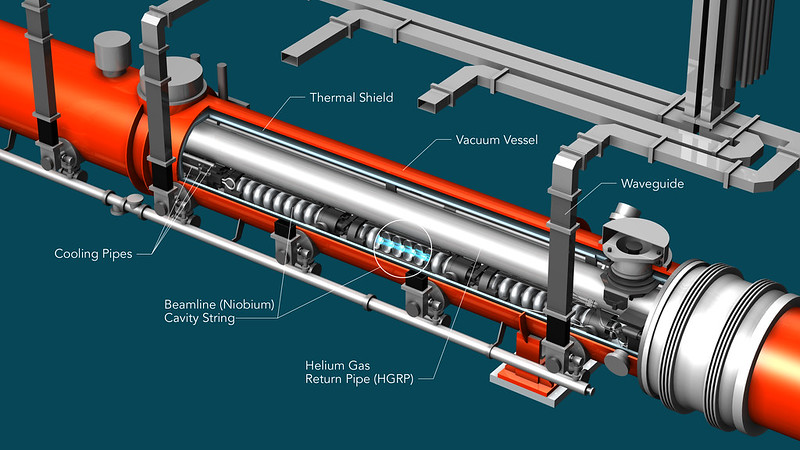}
	\caption{LCLS-II superconducting cryomodule layout.
        Eight niobium cavities are in each cryomodule; three are shown here.
       One of these cryomodules directly follows the components shown in Fig.~\ref{fig:gun}
      and is included in the simulations shown here. The phase and gradient in the first four
     cavities are included as input variables.\label{fig:sc_inj}}
\end{figure*}

The superconducting linear accelerator following the rf gun, buncher, and solenoids will consist of thirty-two cryomodules with
eight rf cavities each, operating at \SI{1.3}{GHz}. See Fig.~\ref{fig:sc_inj} for cryomodule detail.
For the purposes of this paper, only the injector is considered. The injector ends after the first cryomodule, and subsequent hardware is not simulated here.
The beam dynamics (emittance and bunch length) are most sensitive in the first two to three cavities of the first cryomodule.
It is common practice to include phase and gradient in the first four cavities, while keeping the latter four fixed at zero phase and maximum gradient.
Although they may have some small impact on the final results, the reduction in optimization variables is preferred
given the computational resources used/needed to optimize this system.
Note that a zero phase (commonly called ``on crest''), indicates that the cavity phase is timed such that the electron beam rides the `crest' or peak of the rf wave. For a full list of optimization variables and bounds, see Table~\ref{tab:variables}.
\begin{table*}[h]\centering\label{tab:variables}
	\ra{1.3}
	\begin{tabular}{@{}lccll@{}}\toprule
		Variable                   & Minimum & Maximum & Unit  & Scale \\ \midrule
		Solenoid strength(s) & 0.02         &   0.07     & T/m  &   $10^{-3}$\\
		Buncher gradient    & 1.0           & 1.8              & MV/m & 1.0\\
		Buncher phase        & -100        & -10.0           & Degrees &  1.0\\
		Cavity gradient 1-4  & 0.0 		    & 32.0        & MV/m   &  1.0\\
		Cavity phase 1-4     & -40.0       & 40.0        & Degrees &  1.0\\
		\bottomrule
	\end{tabular}
	\caption{Optimization variables and boundaries for the LCLS-II superconducting injector. Scale values are used to adjust the order of variables before passing to optimization algorithms.}
\end{table*}

\subsection{Objectives}
The metrics commonly used to determine photoinjector performance are bunch length ($\sigma_z$) and emittance ($\epsilon$).
The definition of bunch length is the more intuitive of the two, being the longitudinal root mean square size of the electron bunch in physical space.
The emittance is a composite metric that combines momentum and physical space. In this case we consider the emittance in the transverse dimensions,
    \begin{equation}
        \epsilon_x = \sqrt{\langle x_i^2 \rangle \langle p_{x_i}^{2} \rangle - \langle x_i^2 p_{x_i}^{2}  \rangle },
    \end{equation}
where $x_i $ represents the $x$ location of each particle and $p_{x_i}$ represents the momentum in the $x$ direction of each particle~\cite{Wiedemann2015}.
This combination gives feedback on the size of the beam and information on whether it may expand or contract.
Since all electric and magnetic fields in these simulations are symmetric, only the emittance in the $x$ dimension is used as an objective.

The emittance and bunch length are at odds with each other because of space charge forces~\cite{PhysRevAccelBeams.21.010101}.
Based on deficiencies in previous optimizations~\cite{technote,PhysRevAccelBeams.25.013401}, a third objective, energy spread ($dE$), is also included. Optimizations using NSGA-II typically return acceptable emittance and bunch length values but large energy spread distributions. Large amounts of energy spread impact compression of the beam downstream, which results in loss of performance. Large energy spread values have resulted in costly ``tweaking'' of optimization results to adjust the energy spread after optimization. Therefore, the energy spread is added as a third objective in an  attempt to eliminate the need for postoptimization adjustments to the energy spread distribution.
\begin{table*}\centering\label{tab:objectives}
    \ra{1.3}
    \begin{tabular}{@{}lccll@{}}\toprule
        Objective             & Minimum & Maximum & Unit  & Order \\ \midrule
        Emittance           & 0.3        &     6   & $\mu$m  &   $10^{-6}$\\
        Bunch Length    & 0.5           & 4.0              &mm & $10^{-3}$\\
       Energy Spread   & 0.0           & 1.0              & MeV    & $10^{6}$\\
        \bottomrule
    \end{tabular}
    \caption{Objective boundaries for the LCLS-II superconducting injector. These are used to scale the returned objectives to roughly a unit cube. Note that the minimum and maximum are not absolute but rather a general definition of the interesting objective space.}
\end{table*}

\section{Methods and Results}\label{sec:results}
The optimization procedure used in this paper is as follows. A sample of the entire search space is provided to each optimization algorithm. The inputs (design variables) and objectives are scaled the same for each case. For NSGA-II and VTMOP, the algorithm is allowed to evaluate 3,000 simulations. For APOSMM, the 3,000 evaluations were split between five weight values that are representative of the Pareto front. Each weight configuration is given 600 simulation evaluations. In the following sections we compare the samples and the performance of each optimization algorithm  in terms of Pareto optimal points along with performance in convergence.

\subsection{Initial samples}
Because of the number of design variables (twelve) and large search space, initial samples were used as starting points for the optimization algorithms. Two types of sampling methods were used: uniform random sampling and Latin hypercube sampling (LHS)\cite{10.2307/1268522,doi:10.1080/03610928008827996}. In order to show that the sample is not subject to large variability for this problem, five 1,000-point samples with unique seeds were compared for each sampling method. The sample runs are referred to as Runs 1--5 for the remainder of the paper. The objective means from each sample were found to deviate by no more than 15\%. T-tests were done to compare all LHS and uniform samples.
No test resulted in a rejection of the null hypothesis: the samples were found to vary normally, but there were no significant differences between the means of each sample. The similar means and T-test results are expected
and confirm that multiple iterations of uniform random or LHS samples result in similar sample statistics. In the following sections each of the three optimization algorithms is tested with the fifth sample from both the uniform and LHS sets, namely, Run 5. Starting the algorithms from both uniform and LHS will indicate whether the optimization problem is sensitive to the distributions of initial samples.
\begin{figure*}[h]
     \centering
    \includegraphics[width=0.4\linewidth]{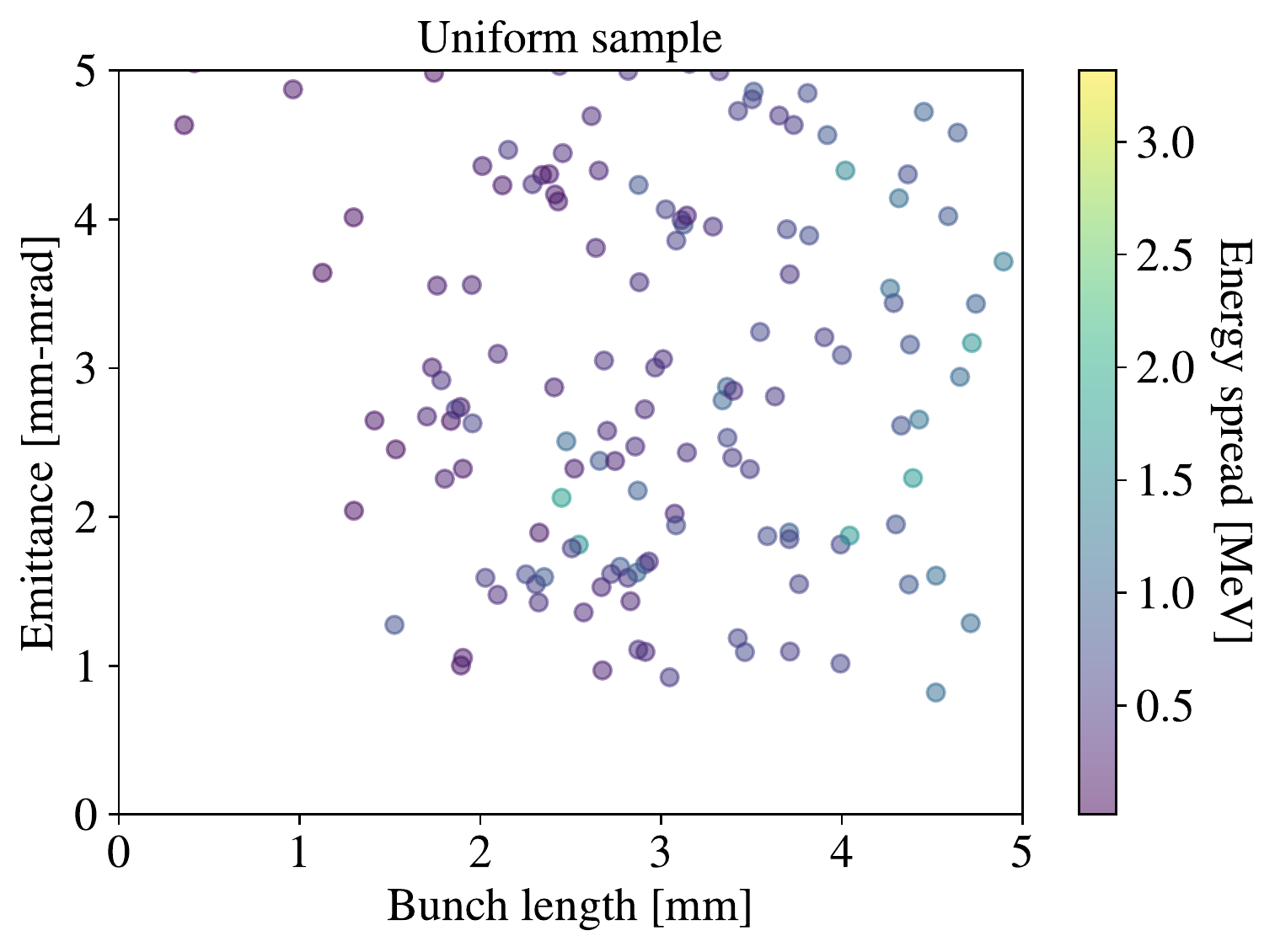}\quad
    \includegraphics[width=0.4\linewidth]{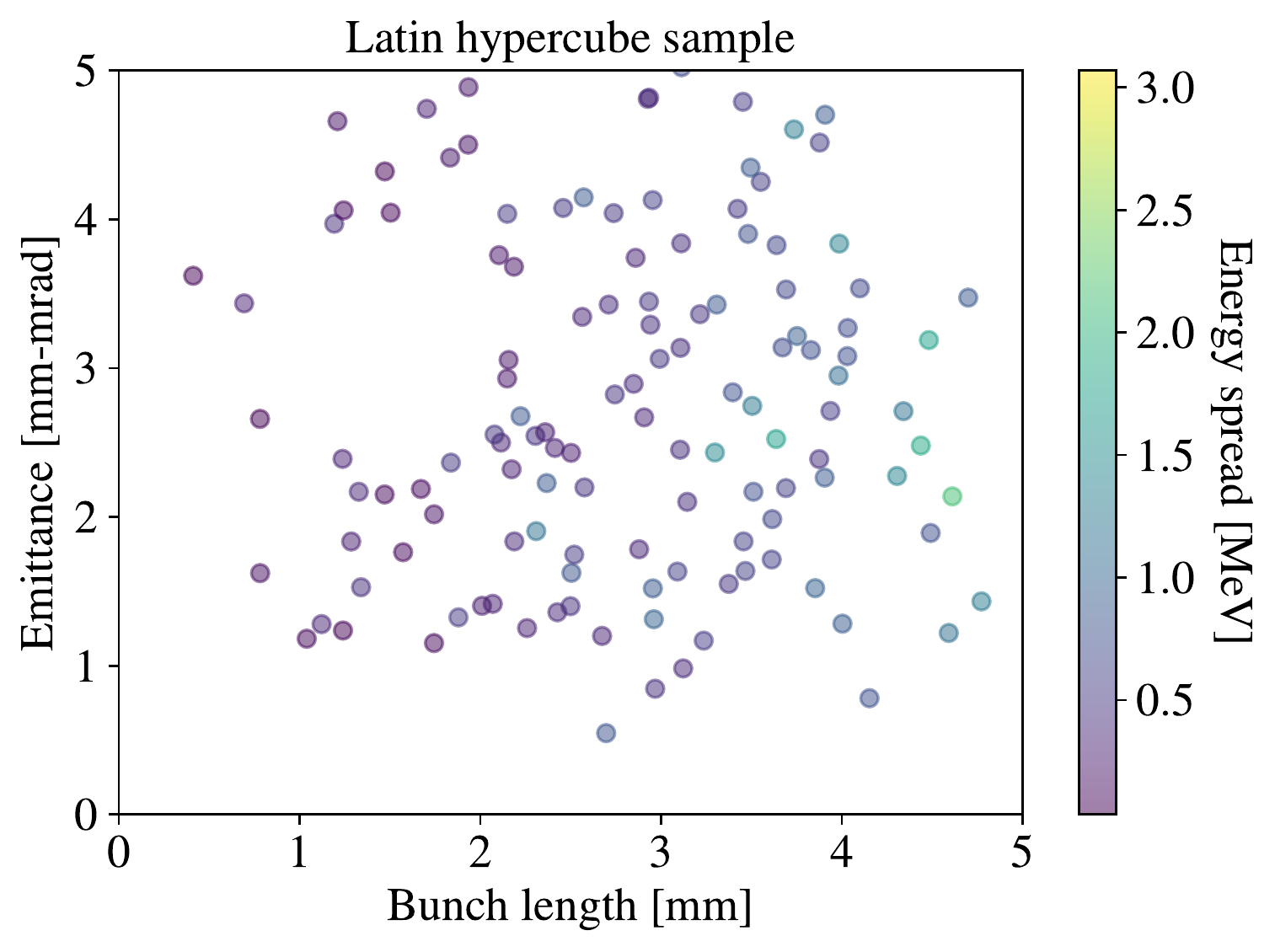}\quad
    \vspace{1em}
    \caption{Uniform and Latin hypercube samples of the LCLS-II injector objective space.
    These samples were given to optimization algorithms as initial points. }
\end{figure*}

\subsection{Penalties}
Simulations are considered viable if two conditions are satisfied. First, did the simulation reach the desired location in the accelerator (\SI{15}{m})? If not, the simulation may have ``hung'' because of a poor combination of input parameters or was terminated because of other factors. This behavior has been observed in past simulations, with the most common cause of hanging simulations being incompatible phases in the rf cavities. Adversarial rf phases can push the electrons backward toward the start of the accelerator, as opposed to the forward direction. This condition traps the simulation in calculations near the cavities with incompatible phases, and the beam bounces back and forth in an oscillatory motion.

The second condition for viability is whether the simulation has lost particles. Particles are lost when the transverse positions ($x_i$) extend beyond the grid boundaries allowed by the input file. This is a conservative number and indicates when particles have traveled beyond the radius of the vacuum dimensions of the accelerator. This condition is especially nefarious because it artificially reduces the emittance, one of the exact objectives we are trying to minimize.

With these metrics defined, consider Run 1 for both the uniform and LH samples. Because the search space is large, many combinations of parameters lead to infeasible solutions, resulting in 243/1,000 uniform and 250/1,000 LH simulations being viable. The large number of invalid points further emphasizes the need to use an initial random sample and penalties for this optimization case. Given the number of failed runs, a penalty is introduced based on the viability metrics described above. The maximum penalty is applied if the simulation lost all particles or failed to reach the desired location. If the simulation completed but lost particles, a penalty proportional to the number of particles lost is applied,
\begin{equation}\label{eq:penalty}
    P_v = \frac{N_{start}-N_{p}}{c}\, ,
\end{equation}
where $P_v$ is the penalty due to viability metrics, $N_{start}$ is the maximum number of particles possible in the simulation, $N_p$ is the number of particles that reached the end of the simulation, and $c$ is a constant applied to scale the maximum penalty value to the order of all objectives. If all particles are lost, the maximum penalty is added to all objectives. If a fraction of the particles is lost, that information is preserved in a scaled penalty that gets worse as more particles are lost.

Domain-specific knowledge gained from previous optimizations is used to formulate one additional penalty. As mentioned in Section~\ref{sec:dvar}, two objective emittance and bunch length optimizations result in the need for energy spread adjustments after the optimization. In past VTMOP runs, adding the third objective, $dE$, resulted in reduced emittance performance. For the VTMOP runs presented here, a penalty is added to all three objectives whenever the emittance exceeds a set threshold. Adding an emittance penalty helps focus the algorithm on lower emittance values. For the LCLS-II, any emittance above $1\,\mu m$ is considered three times larger than the expected baseline performance ($0.3\,\mu m$). In order not to overly constrain the optimization algorithms (another lesson learned from previous work), a conservative threshold of $2\,\mu m$ is used for penalizing emittance values. Evaluations whose emittance is below this threshold are not penalized. The penalty for emittance, $P_e$, is then
\[
P_e=
\begin{cases}
    \epsilon_{x_i}-0.3,& \text{if } \epsilon_{x_i}> 0.3\\
    0,              & \text{otherwise},
\end{cases}
\]
where $\epsilon_{x_i}$ is the scaled emittance. The value of 0.3 approximates the scaled emittance threshold of $2\,\mu m$.
The value 0.3 is subtracted from the scaled emittance to ensure a smooth penalty that increases for larger emittance values. In all optimization runs shown below, each simulation is checked for feasibility; and penalties for failed runs, lost particles, and large emittances are applied if necessary. The total penalty, $P_t$, is then
\begin{equation}
   P_t = P_v + P_e
\end{equation}
and is added to all objectives.

\subsection{Results}
\begin{figure*}[ht!]
    \centering
    \includegraphics[width=0.4\linewidth]{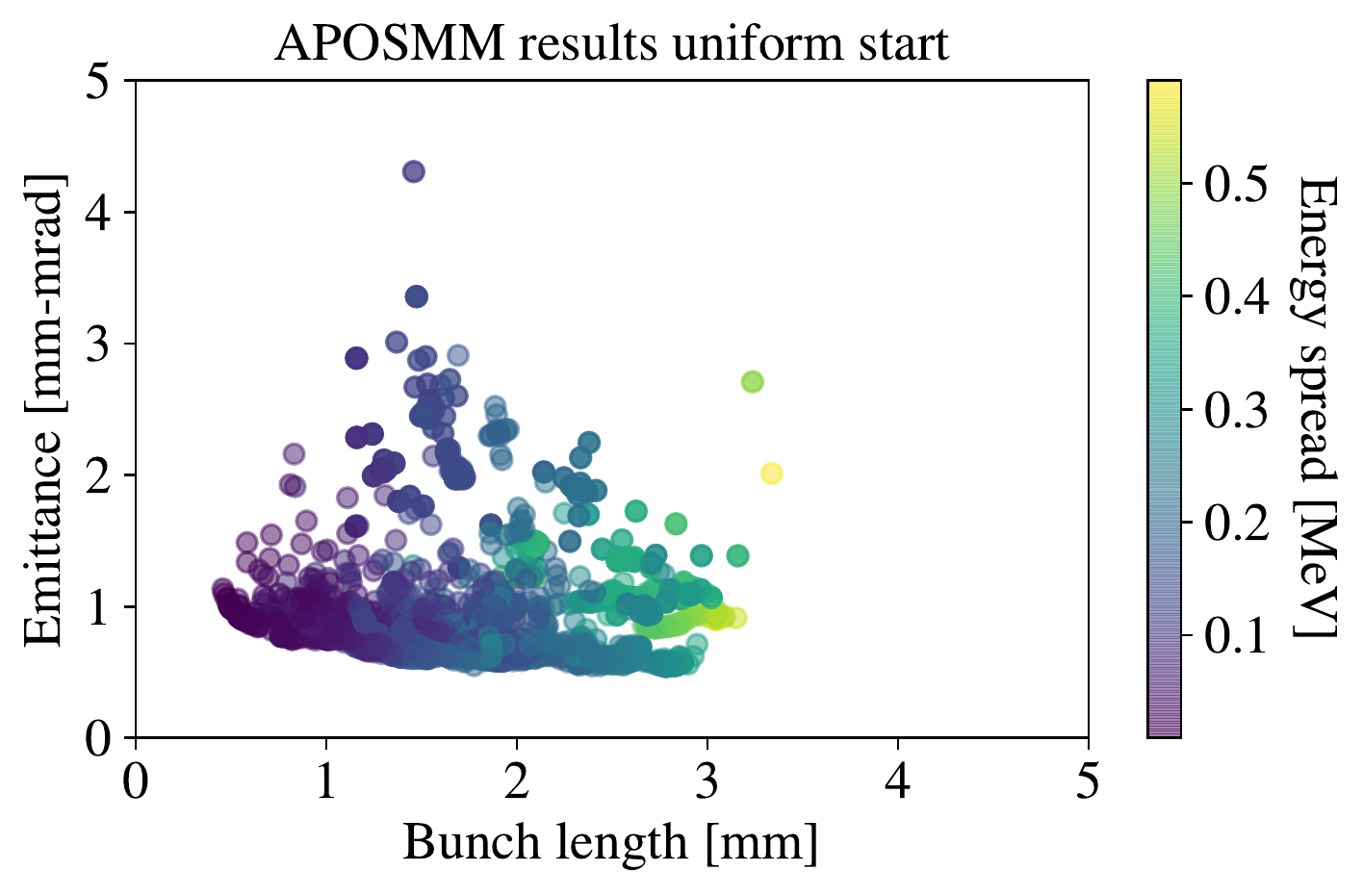}\qquad
    \includegraphics[width=0.4\linewidth]{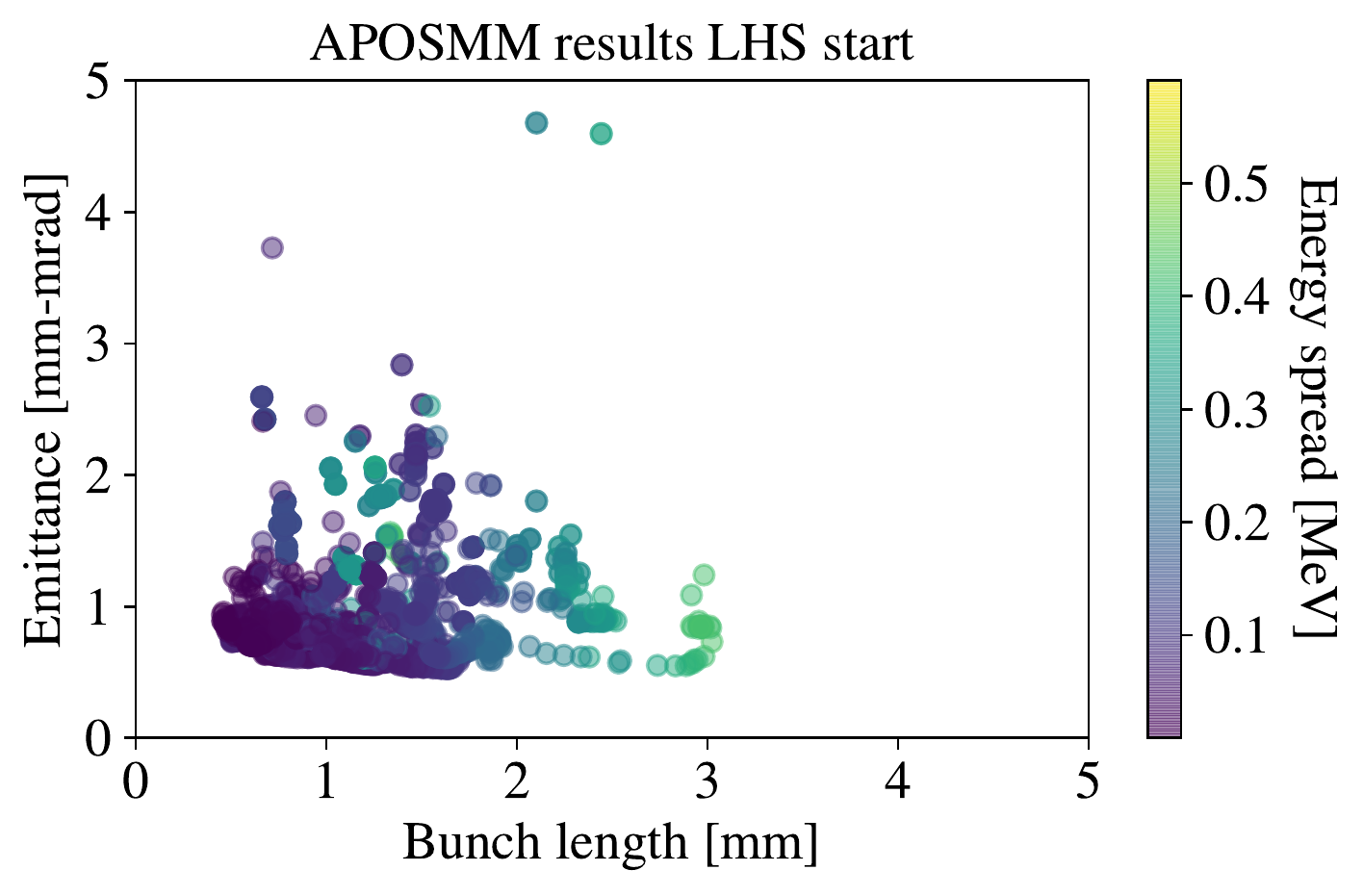}
    \vspace{1em}
    \includegraphics[width=0.4\linewidth]{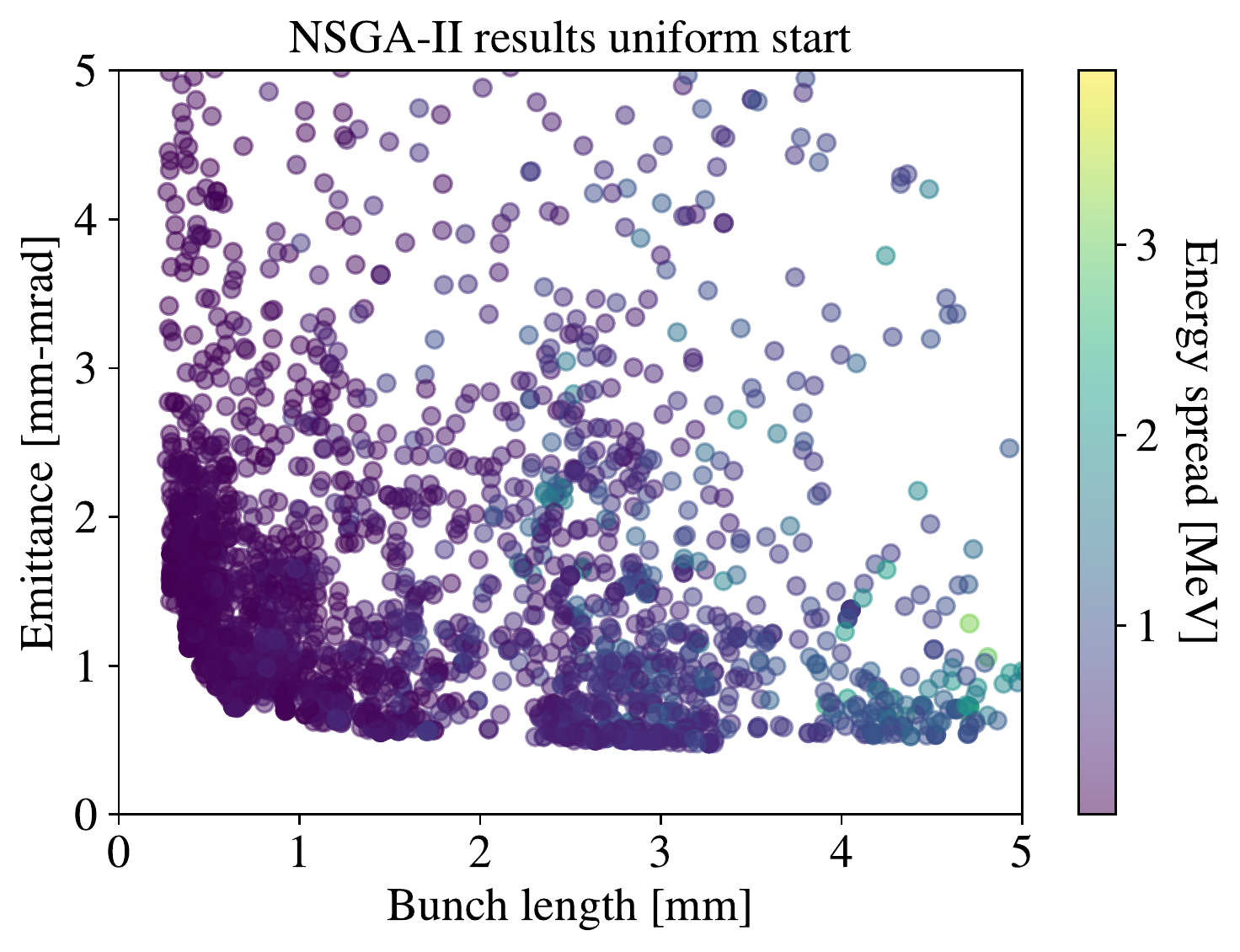}\qquad
    \includegraphics[width=0.4\linewidth]{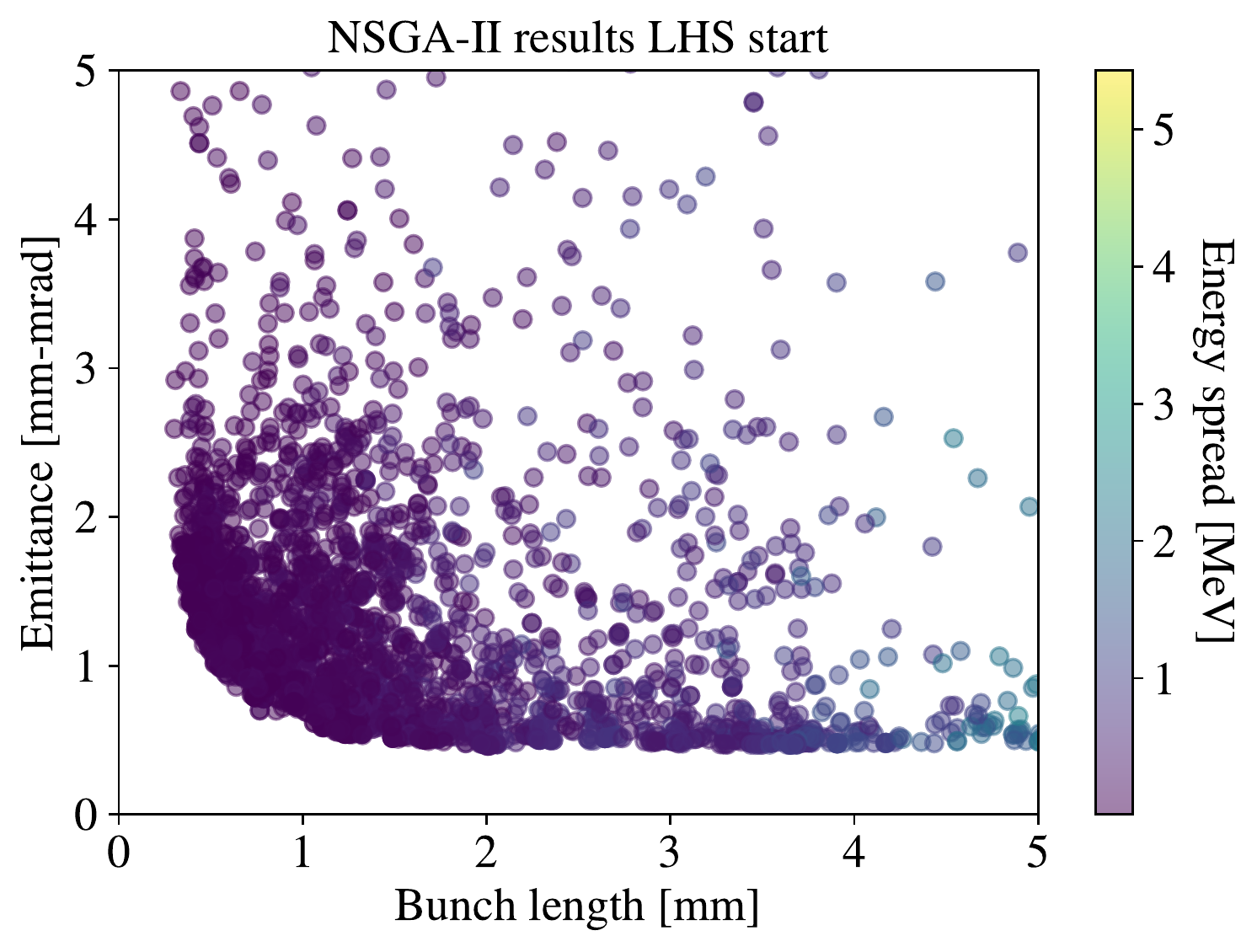}
    \vspace{1em}
    \includegraphics[width=0.4\linewidth]{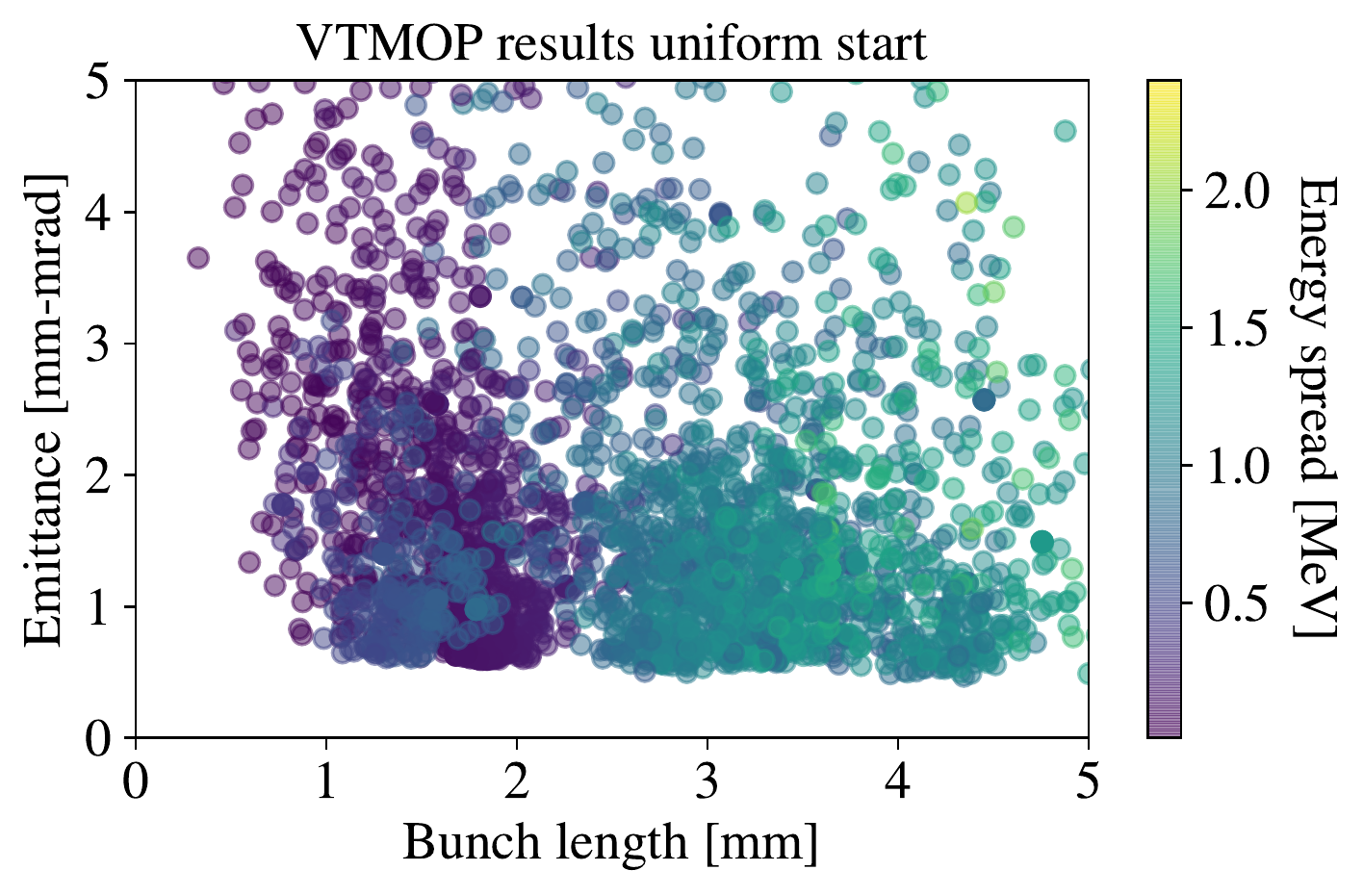}\qquad
    \includegraphics[width=0.4\linewidth]{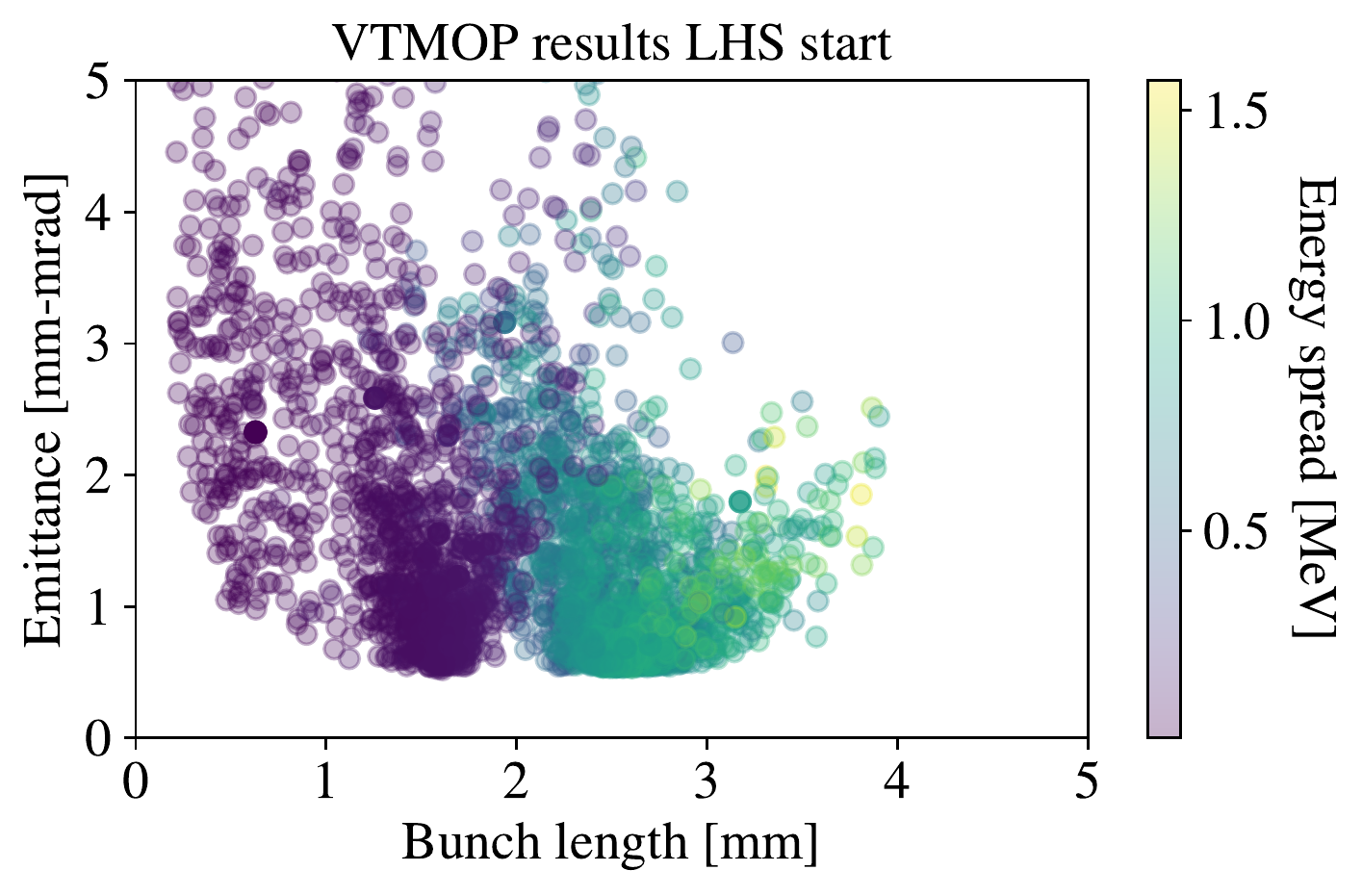}
    \caption{Scatter plots of the samples, APOSMM, NSGA-II, and VTMOP results. The optimization plots do not include the initial sample points. The third objective is plotted with a color bar. Darker regions of the plot indicate a higher density of simulations with low energy spread. Most points shown here are dominated, that is, these are not Pareto optimal plots. This figure indicates the objective values observed by each algorithm.}
    \label{fig:scatter}
\end{figure*}

Using the penalties defined in~\eqref{eq:penalty}, the three algorithms
described in Section~\ref{sec:opt} were used to optimize the LCLS-II injector.
Each algorithm was allowed 3,000 simulation evaluations. 
All simulation evaluations are shown in Fig.~\ref{fig:scatter}.
Plotting all values instead of only nondominated
values gives an understanding of the relative performance of each algorithm.
For example, a high density of points near the Pareto front may
indicate less sensitivity to small variations in magnet or phase parameters.
The opposite
can be said for gaps in the Pareto front. Consider the NSGA-II uniform sample
and VTMOP LHS results around a 2\si{mm} bunch length. Gaps similar to these
have appeared in previous studies~\cite{ml, technote}
and have been sources of uncertainty. It was not clear at the time whether
the gaps arise from the underlying physics being simulated, or because
optimization algorithms were not sufficiently approximating the Pareto front.
Our results show that the gaps resulted from poor
combinations of simulation evaluations and penalties. The penalties and
constraints in previous runs were not sufficient to fill the Pareto front in
the given amount of simulation evaluations. The importance of gaps in the
Pareto front is either exacerbated or alleviated depending on the optimization
goals. APOSMM with an LHS start does not cover the 2\si{mm} bunch length region
well either, but that is not the current operating target at LCLS-II.
Therefore, having a higher density of points around a 1\si{mm} bunch length is
beneficial in this case. If this was the first optimization of LCLS-II, the
NSGA-II LHS results would be the most informative.

\begin{figure*}[h!]
    \centering
    \includegraphics[width=0.9\linewidth]{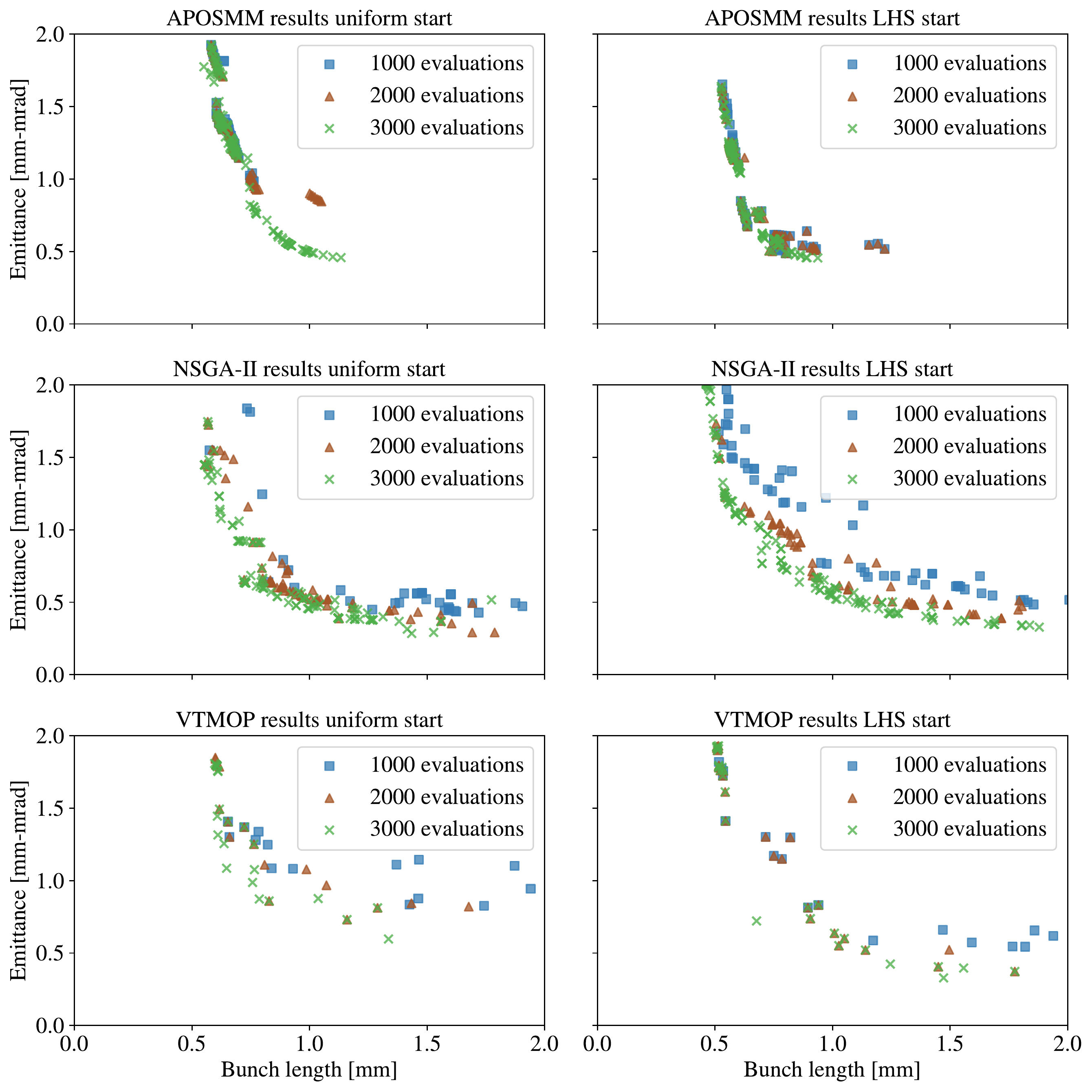}
    \caption{Comparison of Pareto front progression using APOSMM, NSGA-II, and VTMOP. These points are nondominated with respect to all objectives. Each marker type represents a different number of evaluations.\label{fig:pareto_fronts}}
\end{figure*}

Figure~\ref{fig:pareto_fronts} compares the Pareto front for each algorithm
after 1,000, 2,000, and 3,000 evaluations.
The results show a marked difference in optimization progression between the uniform and LH samples.
Further discussion of the causes for these results is presented in Section~\ref{sec:pareto}.
We note the lack of numerical convergence metrics in this section.
As we will discuss in~\ref{sec:pareto}, proper evaluation and comparison
of multiobjective optimization algorithm performance are nuanced.
Because of the difficulty in evaluating multiobjective optimization algorithms,
practitioners often use customized metrics that are specific to the problem at hand.
For this work we have judged that scatter plots of the objective scores
are the best available tool for accelerator physicists
to understand the results of an optimization run.

\subsection{Pareto and convergence comparison}\label{sec:pareto}
It is natural to compare optimization methods based on the objective values observed relative to the effort required by the methods. Because
the cost of evaluating the objective is so large, we desire optimization methods
that can reach a solution in as few simulation evaluations as possible.
In Fig.~\ref{fig:pareto_fronts}, the APOSMM and VTMOP optimizations started
from LHS produce many nondominated points in an approximate Pareto front before
3,000 evaluations. In the case of VTMOP+LHS, several of the Pareto optimal
points are found within 1,000--2,000 evaluations. For both NSGA-II
optimizations, improvements to the Pareto front are made out to 3,000
evaluations. NSGA-II also is the least sensitive to the uniform or LHS start,
with APOSMM and VTMOP performance improving with LHS.

Note that not all portions of the Pareto front are of equal importance to those
running LCLS-II. Interpretation of these results is nuanced, since the absolute
best in any one objective may not be where the accelerator normally runs. In
the case of LCLS-II, the minimum emittance is highly sought and is weighted
heavily, where a bunch length nominally kept at 1 mm is acceptable. A bunch
length of 1 mm is far from the best-achieved minimum of about 0.25 mm in the
VTMOP + LHS optimization. Given that, there are still noted differences in
optimization progression between the algorithms and samples. APOSMM + LHS
achieves the fastest convergence, with most of its progress being made by 500
evaluations. NSGA-II has slow and steady emittance reduction throughout the
3,000 evaluations. VTMOP optimizes emittance the fastest, with stair-step
progress for bunch length and energy spread. The desired bunch length of 1 mm
is reached by 500 evaluations in both VTMOP optimizations.  This is in contrast
to NSGA-II, where the desired bunch length is reached by 500 evaluations but
the emittance (the most important metric) has not reached its minimum until
nearly 2,500 evaluations.

The differences in Pareto convergence can be attributed to the
approach of each algorithm. VTMOP optimizes in one region until a convergence
criterion is met. This results in two dense pockets, as seen in both the uniform
and LHS VTMOP runs. Given more evaluations, VTMOP would further explore the
front. Hyperparameter tuning would impact the density and location of the
explored regions in the VTMOP case. For NSGA-II, the progression out to 3,000
evaluations results from its heuristic approach. Nondominated points were
not found in the initial samples, and NSGA-II relies on statistical
recombination to produce candidate points. Previous evaluations are not
necessarily saved or
considered during each iteration of NSGA-II, meaning progress is often incremental. VTMOP and APOSMM build models that include
previous evaluations and information. This can result in less time spent in
dominated regions of the objective space. We find that APOSMM+LHS is the most
efficient of the three with the most nondominated points found within the first
1,000 evaluations.

While the overall trends in the Pareto curves and progression can be attributed
to the optimization methods, we give one caveat based on the sample. We did not
compare the optimization progression for all ten uniform and LH samples. Note
that any optimization algorithm performance will improve in the case where a
sample point or points are near the Pareto front (which is statistically
possible).

\section{Discussion}\label{sec:discussion}
From this work, several observations and recommendations can be made for the
optimization of LCLS-II and photoinjectors with similar parameters. First, a
random sample is recommended before optimization. This can be either a uniform
or LH sample, with the expectation
that for machines like the LCLS-II, 
an algorithm started with an LHS will reach lower objective values given the
same number of evaluations. The type of algorithm used for optimization should
consider whether the problem requires initial space exploration or whether a
region of interest is known. For example, VTMOP and NSGA-II can be used to
widely explore the Pareto front in cases where there is little prior knowledge
of the system. In contrast, a multistart local method such as APOSMM + BOBYQA
can significantly reduce the number of evaluations needed if there is basic
knowledge of the Pareto front location or if there is interest in specific
regions of the objective space. Very few evaluations are wasted far from the
Pareto front in the APOSMM+BOBYQA cases while still reaching the final Pareto
front. A common justification for avoiding  local optimization methods, such as
BOBYQA, included their tendency to fall into local minima. These results show
that such scenarios can easily be avoided with a multistart (APOSMM) and
multiweight approach.

In contrast, NSGA-II spends many evaluations filling out the entire parameter
space. NSGA-II took more evaluations to reach minimum objective values when
compared to VTMOP or APOSMM, but it did perform better than prior optimization
for this case~\cite{PhysRevAccelBeams.25.013401,technote}. Improved convergence
results may be attributed to the heavy emittance and lost particle
penalization, which was not used in past runs.
In previous optimizations, a lost particle flag was used as a strict
elimination constraint~\cite{technote} instead of a penalty on the objectives.
When the constraint or penalty boundaries were too tight in the
work leading up to~\cite{technote,PhysRevAccelBeams.25.013401}, the NSGA-II
runs would often slow down or fail to converge at all.  It is
hard to emphasize this last point enough.
In many previous optimizations for LCLS-II, NSGA-II runs did not converge
within 3,000 evaluations. This result  clearly indicates that the penalty (all
objectives treated equally) and constraint values were not being used
effectively for photoinjector optimization.
With strong objective penalties, this paper shows convergence can
be reached in many fewer evaluations. Specifically for LCLS-II, an optimization
run can be reduced from 10,000 evaluations or more to 1,000 or 2,000 in the
case of APOSMM+LHS. This is a potential savings of 2,000 CPU hours per
optimization case.

In conclusion, all three methods can achieve the results expected from an optimization of the LCLS-II photoinjector.
The distinction lies in three key areas. What sample is used? What
penalties are used? And what is the goal of the optimization study? For
LCLS-II, LHS samples improve the performance of all three methods. The
objective space is fairly well known at this time, so strong penalties on
emittance were implemented. This sped up optimization convergence. If the goal
of an optimization run is to explore a specific region of the objective space,
time and CPU hours can be saved by using a multistart local algorithm (e.g.,
APOSMM+BOBYQA) or a model-based method (e.g., VTMOP). Needing fewer
evaluations for targeted optimizations may enable higher-fidelity simulations
in narrow regions of the objective space. If initial exploration is the goal,
NSGA-II can evenly explore the Pareto front given enough evaluations.

We note that our suggestion to penalize large emittance scores
may be difficult to implement without a priori knowledge of the
range of attainable emittance scores.
Additionally, one may be able to further improve results by penalizing
other uninteresting regions of the Pareto front.
In future work, then, one might consider
{\it interactive multiobjective optimization} algorithms, which integrate
a decision-maker's updated preferences into the optimization loop \cite{afsar2021assessing}.

\section{Code Implementations}\label{sec:code}
This section details the code implementation and resources used.
Preference was given to open-source and freely available packages and simulation codes.
All software used to simulate the accelerator and run the optimizations are version controlled with Git and available on GitHub or GitLab.

\subsection{Simulation model}
The particle-in-cell simulation code OPAL~\cite{opal} is used to model the accelerator.
This is an open-source C++ code available on GitLab:
\url{https://gitlab.psi.ch/OPAL/src/-/wikis/home}.
Two flavors are available: OPAL-t and OPAL-cycl.
The latter is used for simulating cyclotrons, and in this paper OPAL-t was used.
A convergence study was done around a previously found optimal from the simulation
code ASTRA~\cite{astra}.
The final grid size used for these runs was $32^3$ with 50,000 particles.
Two time steps were used: $10^{-13}$ to \SI{0.2}{m} and $10^{-12}$ to \SI{15}{m}.
Magnetic and electric field maps were included in the simulations along with space charge.
Initial laser distributions were simulated with the Python package Distgen:
\url{https://github.com/ColwynGulliford/distgen}.
Results from Distgen were then supplied to OPAL, as the initial distribution applied to the cathode.

\subsection{libEnsemble}
The open-source Python library  libEnsemble was used to manage all optimization
runs and simulations~\cite{libEnsemble}.
libEnsemble is designed to coordinate the concurrent evaluation of
dynamic ensembles of calculations. The library is developed to use massively
parallel resources to accelerate the solution of design, decision, and
inference problems. When the beamline simulation reaches a point where it
no longer speeds up when given additional parallel computing resources,
libEnsemble allows us to more efficiently utilize such resources by
coordinating concurrent simulation evaluations at different parameter values.
libEnsemble employs a manager/worker scheme that can run on various
communication media, and it supports
executables such as OPAL.

libEnsemble enables the workflow to be composed of a generator script expressing
the optimization method and a simulator script to manage the simulations
and kill them if a timeout is reached. The simulator script uses a
libEnsemble-provided executor interface that can detect and manage resources and system
MPI runners, as well as run, poll, and kill simulations, and is thus highly
portable between platforms. libEnsemble enables the placement of multiple concurrent
OPAL runs on each node by specifying the number of workers required.

\subsection{Computational resources}
Preliminary runs and tests were performed on Bebop at Argonne National Laboratory. Bebop consists of Broadwell (36 cores) or Knights Landing (64 cores) compute nodes.  Final runs were performed on Bubble at SLAC, which consists of Broadwell (36 cores) nodes. A typical four-core MPI OPAL simulation takes approximately 4.2 minutes on a Broadwell compute node. All sampling evaluations and APOSMM, NSGA-II, and VTMOP runs were performed on either one or two nodes, using 36 or 72 cores, respectively. These are modest resource scales and intended to match what is readily available at most laboratories and universities.

\section*{Acknowledgments}
This material is based upon work supported by the U.S. Department of Energy,
Office of Science, under contract numbers DE-AC02-76SF00515 and
DE-AC02-06CH11357.
This work was supported in part by the U.S.~Department of Energy, Office of
Science, Office of Advanced Scientific Computing Research and Office of
High-Energy Physics, Scientific Discovery through Advanced Computing (SciDAC)
Program through the FASTMath Institute and the ComPASS-4 Project under Contract
No.~DE-AC02-06CH11357.
We gratefully acknowledge the computing resources provided on Bebop and Bubble,
high-performance computing clusters operated by the Laboratory Computing
Resource Center at Argonne National Laboratory, and the Scientific Data
Facility at SLAC.
Special thanks to Gregory Stewart and the METSD department at SLAC for the CAD
and layout drawings.

\printbibliography

\vfill
\framebox{\parbox{.90\linewidth}{\scriptsize The submitted manuscript has been created by
        UChicago Argonne, LLC, Operator of Argonne National Laboratory (``Argonne'').
        Argonne, a U.S.\ Department of Energy Office of Science laboratory, is operated
        under Contract No.\ DE-AC02-06CH11357.  The U.S.\ Government retains for itself,
        and others acting on its behalf, a paid-up nonexclusive, irrevocable worldwide
        license in said article to reproduce, prepare derivative works, distribute
        copies to the public, and perform publicly and display publicly, by or on
        behalf of the Government.  The Department of Energy will provide public access
        to these results of federally sponsored research in accordance with the DOE
        Public Access Plan \url{http://energy.gov/downloads/doe-public-access-plan}.}}

\end{document}